\documentclass{article}
\usepackage[T1]{fontenc}
\usepackage[utf8]{inputenc}
\usepackage{ismir} 
\usepackage{amsmath,cite,url}
\usepackage{graphicx}
\usepackage{color}
\usepackage{paralist}

\makeatletter
\define@key{Gin}{alt}{}
\makeatother

\title{STASE: A Spatialized Text-to-Audio Synthesis Engine for Music Generation}

\multauthor
  {Tutti Chi$^1$ \hspace{1cm} Letian Gao$^2$ \hspace{1cm} Yixiao Zhang$^3$}
  {{\bf $^1$ University of Chinese Academy of Sciences, China}\\
   {\bf $^2$ Tsinghua University, China}\\
   {\bf $^3$ Centre for Digital Music (C4DM), Queen Mary University of London, UK}\\
   {\tt\small chengtopia@outlook.com, glt23@mails.tsinghua.edu.cn, yixiao.zhang@qmul.ac.uk}
  }
\def\authorname{T. Chi, L. Gao, and Y. Zhang}

\begin{document}

\maketitle

\begin{abstract}
While many text-to-audio systems produce monophonic or fixed-stereo outputs, generating audio with user-defined spatial properties remains a challenge. Existing deep learning-based spatialization methods often rely on latent-space manipulations, which can limit direct control over psychoacoustic parameters critical to spatial perception. To address this, we introduce STASE, a system that leverages a Large Language Model (LLM) as an agent to interpret spatial cues from text. A key feature of STASE is the decoupling of semantic interpretation from a separate, deterministic signal-processing-based spatial rendering engine, which facilitates interpretable and user-controllable spatial reasoning. The LLM processes prompts through two main pathways: (i) Description Prompts, for direct mapping of explicit spatial information (e.g., ``place the lead guitar at $45^{\circ}$ azimuth, 10 m distance''), and (ii) Abstract Prompts, where a Retrieval-Augmented Generation (RAG) module retrieves relevant spatial templates to inform the rendering. This paper details the STASE workflow, discusses implementation considerations, and highlights current challenges in evaluating generative spatial audio.
\end{abstract}

\section{Introduction}\label{sec:introduction}
The landscape of AI-powered music generation has advanced rapidly, moving from general-purpose frameworks towards precision-oriented paradigms \cite{wang2024review, copet2023simple, agostinelli2023musiclm}. In parallel, immersive audio is gaining traction across industries, fueling demand for spatially enhanced auditory experiences \cite{potter2022relative,immohr2023proof}. However, current spatial audio synthesis workflows---such as Dolby Atmos and Apple renderers---have been reported by music producers to provide limited operational controllability \cite{dewey2024practitioners}.

Most existing deep learning-based spatialization methods operate solely on audio inputs, without leveraging direct textual descriptions \cite{leng2022binauralgrad, parida2022beyond, liu2023audioldm}. While latent-space manipulation techniques show potential \cite{li2024tas,heydari2024immersediffusion, sun2024both}, their black-box nature may result in information loss and limit precise control over psychoacoustic parameters critical to spatial perception \cite{radford2021learning,abdal2019image2stylegan,rombach2022high}. Unlike features such as melody or emotion, spatial information can be explicitly modeled and manipulated via signal processing.

\begin{figure}[ht]
    \centering
    \includegraphics[width=\linewidth]{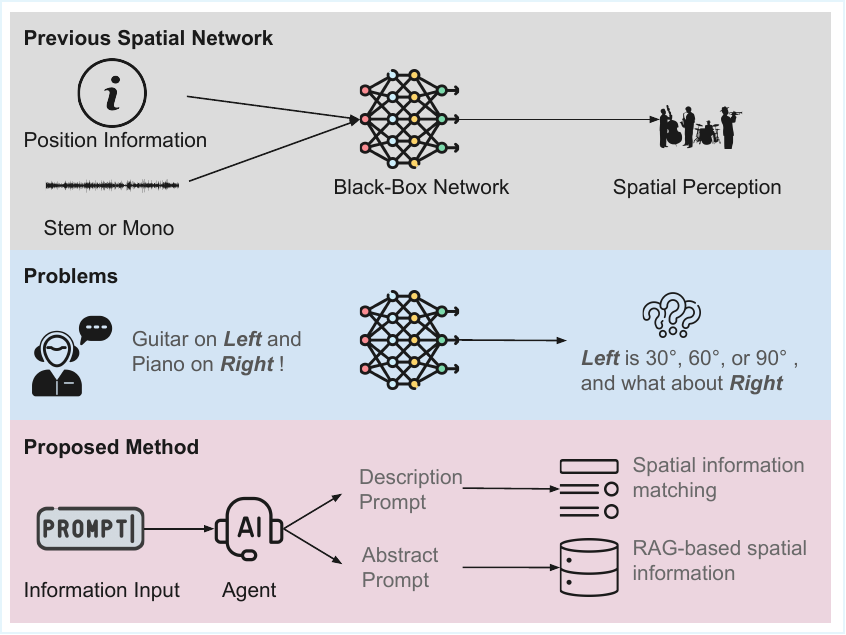}
\caption{High-level comparison between latent-space spatialization pipelines and the proposed agentic STASE pipeline. STASE decouples prompt interpretation (LLM + RAG) from a deterministic signal-processing-based renderer to improve controllability and interpretability.}
    \label{fig:methods}
\end{figure}

Leveraging established principles of binaural hearing---including interaural time difference (ITD), interaural level difference (ILD) \cite{rayleigh1875our,blauert1997spatial}, and dynamic cues \cite{wallach1940role}---we introduce \textbf{STASE}, a hybrid neuro-symbolic framework for generating spatially dynamic music from natural language prompts. As illustrated in Fig.~\ref{fig:methods}, prior spatial networks directly map position information and audio signals (stem or mono) to spatial perception via a black-box model, which can have difficulty interpreting spatial cues from descriptive language and provides limited controllability. In contrast, STASE integrates an LLM as an agent to process the input prompt and route it into one of two pathways: (\emph{i}) \emph{Description Prompts}, enabling direct spatial information matching when precise positional details are given; and (\emph{ii}) \emph{Abstract Prompts}, where a Retrieval-Augmented Generation (RAG) module retrieves relevant spatial knowledge before rendering. This decoupling of semantic interpretation from deterministic signal-processing-based spatial rendering supports interpretable and user-controllable spatial reasoning.

Our modular architecture combines the LLM-based prompt interpreter, a music generation module for content creation, optional source separation or monaural instrument inputs, and a dedicated spatialization engine driven by LLM-derived parameters. In our current implementation, stems are generated via a music synthesis model and source separation is not required. This modularity allows components to be independently replaced or fine-tuned, supporting both novice-friendly presets (e.g., fixed studio arrangements) and expert-level customized synthesis.

\section{Related Work}\label{sec:related}
\subsection{Spatial Audio Perception and Production}\label{sec:hearing}
Research on spatial auditory perception, starting with Rayleigh's foundational work \cite{rayleigh1875our}, has highlighted the roles of pinna filtering, Head-Related Transfer Functions (HRTFs), and ITDs in sound localization. Later studies demonstrated how dynamic head movements \cite{wallach1940role} and individualized HRTFs \cite{mendoncca2012improvement} improve both localization precision and subjective immersion. Despite these advancements, a study involving music producers noted that current spatial audio tools often lack flexibility and face challenges with playback consistency across different devices \cite{dewey2024practitioners}.

\subsection{Intelligent Music Generation Systems}
The field of music AI has seen rapid progress, particularly with the emergence of LLMs, leading to systems such as MusicGen \cite{copet2023simple} and MusicLM \cite{agostinelli2023musiclm}. These models have been further refined for specific tasks such as instruction following \cite{zhang2024instruct} or temporal control \cite{lan2024musicongen}. While many AI models now generate stereo audio (e.g., MusicGen \cite{copet2023simple}, Jen-1 \cite{yao2023jen}, Stable Audio \cite{evans2024fast}), they generally lack the precise spatial control needed for detailed spatial rendering and immersive experiences.

\subsection{Spatial Audio Generation Techniques}
Stereo panning is a fundamental music production technique, deeply integrated into Digital Audio Workstations (DAWs) and primarily based on amplitude panning using principles like the "sine-cosine" pan law \cite{pulkki2001spatial}. Beyond panning, the broader field of spatial sound synthesis and recording is an active research area. Recent neural network methods have emerged, such as BinauralGrad \cite{leng2022binauralgrad} for binaural signal conversion and methods for learning spatial cues from video \cite{parida2022beyond}. Other approaches, like AudioLDM \cite{liu2023audioldm}, manipulate audio effects via latent spaces, while TAS \cite{li2024tas} spatializes monaural audio through similar latent manipulations. More recently, ImmerseDiffusion \cite{heydari2024immersediffusion} generates Ambisonics from text, and Sun et al. \cite{sun2024both} proposed a language-driven stereophonic audio generation framework. Our work builds upon these, specifically focusing on achieving precise text-to-spatial-audio control for music.

\section{Methodology}\label{sec:methodology}

STASE is an LLM-driven spatial audio synthesis framework designed to generate musical compositions with user-specified spatial attributes from natural language prompts. The system architecture transforms textual inputs into a fully rendered, multi-track spatial audio mix through a modular pipeline.

\begin{figure*}[htbp]
\centering
\includegraphics[width=1.0\linewidth]{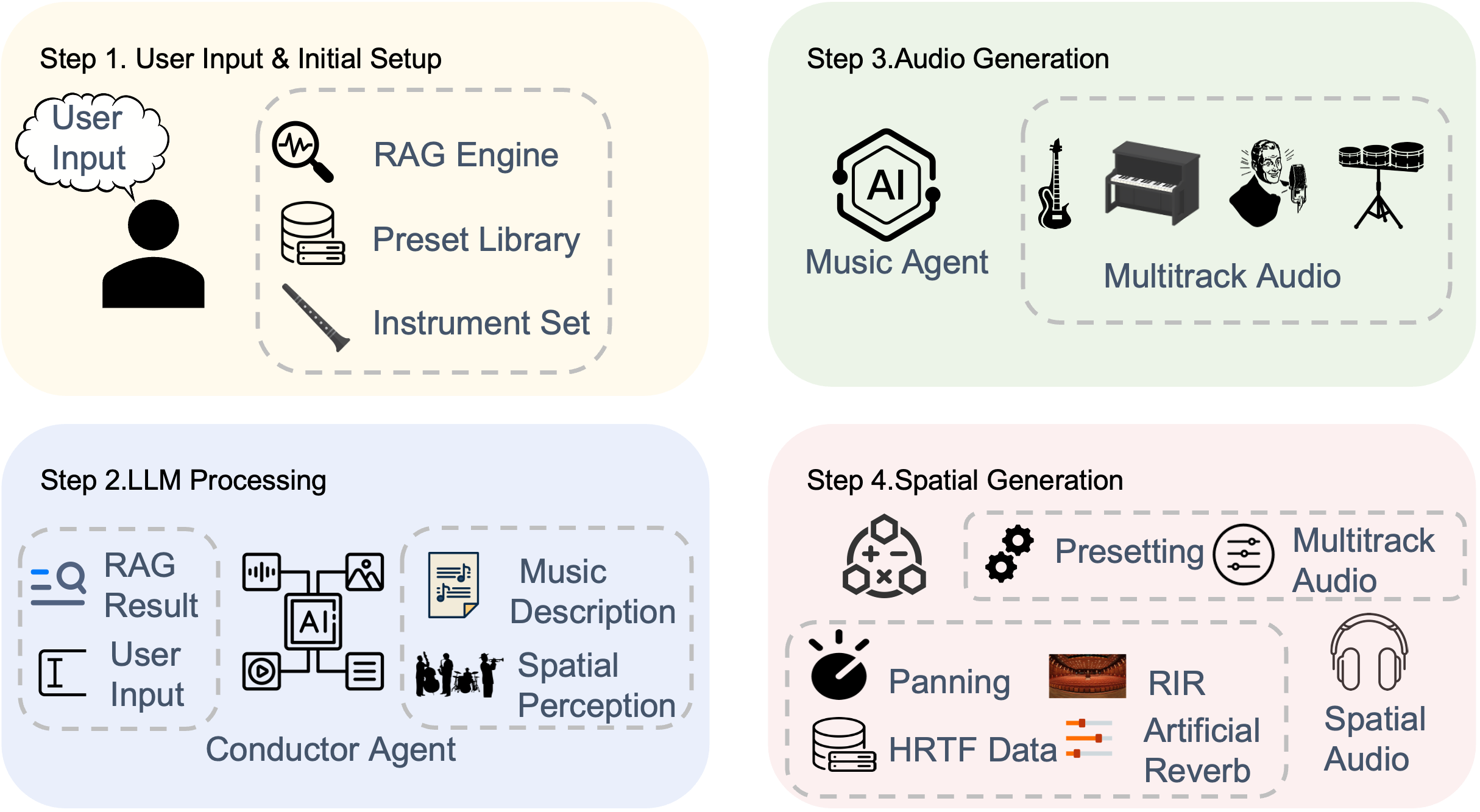}
\caption{STASE workflow: prompts are fused with template knowledge (RAG), transformed by a Conductor Agent into a structured plan (music description, spatial map, mix notes), synthesized into stems, and rendered by a deterministic signal-processing chain (per-source localization uses one of: panning, ITD/ILD, or HRTF; plus reverb/RIR). The agent routes inputs via Description vs. Abstract pathways.}
\label{fig:pipeline}
\end{figure*}

As depicted in Figure \ref{fig:pipeline}, STASE operates in four sequential stages. First, it ingests a free-form natural language prompt. A RAG module searches a preset library for semantically relevant spatial configurations to inform the generation process. Second, the \emph{Conductor Agent}—a core reasoning module—fuses the raw prompt with the retrieved presets. Driven by an LLM, this agent outputs a structured plan: standardized music descriptions, exact spatial parameters (azimuth, distance, etc.), and concise mixing directives. Third, a music generation model synthesizes individual audio tracks (stems) based on the Conductor Agent's descriptions. Finally, the spatial renderer applies the agent-derived parameters—using one of panning, ITD/ILD, or HRTF for localization, plus reverberation—to the stems, yielding the finished spatial mix.

\subsection{Prompt Interpretation and Adaptation}

The essence of STASE lies in translating diverse natural-language prompts into precise, actionable spatial parameters via an LLM. As illustrated in Figure \ref{fig:pipeline}, the process starts when the user supplies a description of the desired sonic scene, optionally augmented by presets, instrument sets, and the RAG engine. The LLM, acting as a \emph{Conductor Agent}, processes both the raw prompt and any RAG-retrieved results to produce a music description and a structured spatial perception map. If the prompt contains explicit spatial cues (e.g., ``place the lead guitar at $45^{\circ}$ azimuth, 10\,m distance''), the LLM directly parses these into quantifiable parameters such as azimuth, elevation, and distance; if the prompt is more abstract (e.g., ``a grand orchestral arrangement''), the LLM queries the RAG module to retrieve semantically relevant spatial templates (e.g., ``symphony orchestra stage setup'') and maps these to default parameter sets. The music description is then forwarded to the \emph{Music Agent} to generate multitrack audio, ensuring that spatial placement decisions are tied to clearly defined sound sources. Finally, the spatialization engine applies either user-specified coordinates or template-driven defaults to the multitrack audio, incorporating techniques such as panning, HRTF-based binaural rendering, Room Impulse Response (RIR) convolution, and artificial reverberation to produce the final spatial audio. To accommodate varying user expertise and prompt specificity, STASE selects the spatialization approach:
\begin{itemize}
    \item \textbf{Precise Spatialization:} When the input prompt contains explicit spatial cues (e.g., "place the lead guitar at 45° azimuth, 10 meters distance"), the LLM directly extracts and applies these values for accurate source positioning.
    \item \textbf{Templated Spatialization:} For more abstract or less precise prompts (e.g., "a grand orchestral arrangement"), the LLM utilizes a RAG approach. It retrieves pre-defined spatial templates (e.g., "symphony orchestra stage setup") semantically closest to the input, providing default spatial parameters for coherent layouts.
\end{itemize}

\subsection{Deterministic Signal-Processing-Based Spatial Rendering}

STASE employs a flexible, deterministic signal-processing approach for spatial rendering (as opposed to latent-space manipulation). Our spatialization module supports established signal-processing methods for sound source localization and environmental acoustic simulation:

\subsubsection{Sound Source Localization}

We support three mutually exclusive localization modes: (i) stereo amplitude panning for lateral placement and non-binaural playback; (ii) analytic ITD/ILD rendering (without HRTF) for controlled cue manipulation; and (iii) HRTF convolution for full three-dimensional binaural rendering. The Conductor Agent selects one mode per source based on the requested coordinates and the target output format. When HRTF is used, additional panning or explicit ITD/ILD processing is disabled to avoid double-counting cues. In non-HRTF modes, ITD is realized by fractional-delay lines aligned with head-width approximations and ILD by frequency-dependent gain shaping.

\subsubsection{Reverberation}

As with localization, reverberation enriches perceived space. Based on the input style, the system can use parameterized algorithmic reverberators for fine-grained control or directly convolve with Room Impulse Responses (RIRs) for specific acoustic environments or stylistic matching. RIR selection can be initialized from the template retrieved by RAG and refined by the Conductor Agent's mix notes.

\section{Implementation Details}\label{sec:implementation}

\subsection{Resources}\label{sec:resources}

We describe resources used in our implementation, including acoustic databases and predefined spatial configurations.

\subsubsection{RIR Database}
For environmental acoustic simulation, we employed single-channel RIRs extracted from established multi-channel datasets including the dEchorate database \cite{dechorate2021} and selected measurements from the OpenAIR library \cite{openair2009}. Our curated RIR collection includes ten distinct acoustic environments designed to match our spatial configuration templates: (1) large concert halls with natural reverberation for classical orchestras, (2) intimate studio rooms for jazz ensembles, (3) dry recording studios for controlled rock band setups, (4) small chambers for intimate music arrangements, (5) medium-sized venues for electronic performances, (6) churches with extended reverb for choir formations, (7) recital halls for solo performances, (8) acoustically diverse spaces for world music, (9) professional recording environments, and (10) simulated outdoor spaces with minimal reflections for festival configurations. Each single-channel RIR is applied to individual audio stems during the spatial rendering process, enabling the system to match textual descriptions of acoustic environments with appropriate reverberation characteristics that complement the corresponding spatial arrangement template.

\subsubsection{Spatial Configuration Templates}
To systematically evaluate spatial placement accuracy, we manually defined ten distinct spatial configuration templates corresponding to common musical performance scenarios: (1) \emph{Classical Orchestra} - traditional symphonic layout with woodwinds front, brass middle, and strings distributed; (2) \emph{Jazz Ensemble} - intimate small group arrangement; (3) \emph{Rock Band} - conventional stage setup with drums center-back; (4) \emph{Chamber Music} - close-proximity classical arrangement; (5) \emph{Electronic/DJ Setup} - electronic music performance configuration; (6) \emph{Choir Formation} - vocal ensemble positioning; (7) \emph{Solo Performance} - single instrument with accompaniment; (8) \emph{World Music Ensemble} - diverse cultural instrument arrangements; (9) \emph{Studio Recording} - controlled studio environment layout; and (10) \emph{Outdoor Festival} - large-scale outdoor performance setup. Each template specifies precise azimuthal positions, elevation angles, and distance parameters for up to 6 simultaneous sources, providing standardized reference points for evaluation.

\subsubsection{HRTF Implementation}
For binaural spatialization, we employed the KEMAR HRTF database \cite{gardner1995hrtf}. 

\subsection{LLM and Prompting Details}
We use an instruction-tuned, open-weight LLM in the 7--13B parameter range with deterministic decoding (temperature $=0$, top-$p=1$) and a constrained, schema-guided output format. Few-shot exemplars cover both Description and Abstract pathways. We provide model names and prompt templates with the supplementary materials.

\subsection{Music Generation Module}
Our prototype integrates an off-the-shelf text-to-music synthesis system to produce up to 2--6 stems depending on the prompt (e.g., drums, bass, guitar, keys, lead, pads). The module can be substituted or bypassed without changing the rest of the pipeline.

\subsection{User-Provided Stems}
Beyond generation, STASE accepts user-provided monaural stems. The user supplies instrument labels (or lets the Agent infer them), and the spatial renderer applies the same plan to the uploaded stems, enabling practical workflows in DAW-centric production.

\subsection{Reproducibility Notes}
We release the template bank, RIR list, parsing code, and prompts used for the Conductor Agent, together with random seeds and decoding settings. This enables step-by-step reproduction of routing decisions and rendering given fixed inputs.

\section{Results}\label{sec:results}
Audio demonstrations are available on the project page: \url{https://chengtopia.github.io/STASE.github.io/}. We include multiple RAG-driven generations and audio samples to facilitate subjective evaluation of spatialization quality and controllability.

\section{Discussion}\label{sec:discussion}
Evaluating text-driven spatial audio remains challenging due to the lack of standardized metrics. ITD and ILD are effective for single-source azimuthal accuracy (e.g., 30° vs 60°), but in multi-source mixes their cues interact and are hard to isolate; they also fail to capture full 3D or perceptual quality. In complex arrangements, overlap and reverberation further confound per-source analysis. Conventional semantic metrics (e.g., CLAP, T5/KL) measure content alignment yet are largely insensitive to fine-grained spatial instructions. We therefore recommend paired objective proxies together with controlled listening tests tailored to spatial attributes.

\section{Conclusion}\label{sec:conclusion}
We presented STASE, an agentic framework that interprets prompts with an LLM and renders spatial mixes using a deterministic signal-processing chain. Decoupling semantics from rendering improves controllability and interpretability, and the modular design supports swapping LLMs, music models, and spatializers. The main limitation is reliance on separated monaural stems; performance can degrade on dense or reverberant mixtures. Future work includes standardized objective/subjective evaluation for spatial attributes and extending the interaction model to mixed monaural inputs and broader production workflows.

\bibliography{ISMIRtemplate}

\end{document}